\documentstyle[12pt]{article}

\begin{document}

\pagestyle{empty}

\hfill hep-th/9612119

\hfill CERN-TH/96-351

\hfill USC-96/HEP-B7

\hfill LPT ENS 96/70

\bigskip \bigskip \bigskip 

\begin{center}
{\LARGE A\ NEW\ SUPERSYMMETRY}
{\LARGE \ } 
\bigskip \bigskip \bigskip \\[0pt]

{\Large Itzhak Bars }{$^a$}  
{\Large and Costas Kounnas }{$^b$} {\Large \ \bigskip \bigskip \\[0pt]}

{\large TH Division, CERN, CH-1211 Geneva 23, Switzerland}

\medskip \bigskip \bigskip \bigskip \bigskip

{\bf ABSTRACT}
\end{center}

\noindent
{We propose a new supersymmetry in field theory that generalizes standard
supersymmetry and we construct field theoretic models that provide some of
its representations. This symmetry combines a finite number of ordinary four
dimensional supersymmetry multiplets into a single multiplet with a new type
of Kaluza-Klein embedding in higher dimensions. We suggest that this
mechanism may have phenomenological applications in understanding family
unification. The algebraic structure, which has a flavor of W-algebras, is
directly motivated by S-theory and its application in black holes. We show
connections to previous proposals in the literature for 12 dimensional
supergravity, Yang-Mills, (2,1) heterotic superstrings and Matrix models
that attempt to capture part of the secret theory behind string theory.
}

\bigskip \bigskip

\noindent CERN-TH/96-351

\noindent {December 1996}

\vskip 1cm
\vfill
\hrule width 6.7cm \vskip 2mm

{$^a$ {{\small On sabbatical leave from the
Department of Physics and Astronomy, University of Southern California, Los
Angeles, CA 90089-0484, USA.}}}

{$^b$ {{\small On leave from 
Ecole Normale Sup\'erieure, 24 rue Lhomond, F-75231, Paris, Cedex 05, FRANCE. }}}

\vfill\eject

\setcounter{page}1\pagestyle{plain}

\section{New supersymmetry}

The usual $N=1$ superalgebra in four dimensions is 
\begin{equation}
\left\{ Q_\alpha ,\bar{Q}_{\dot{\beta}}\right\} =\sigma _{\alpha \dot{\beta}
}^\mu \,p_\mu \,\,,
\end{equation}
where $\alpha ,\dot{\beta}$ are Weyl spinor indices corresponding to $\left(
1/2,0\right) $ and $\left( 0,1/2\right) $ representations of the Lorentz
group $SO(3,1)$. The simplest example of the new superalgebras that we will
discuss in this paper is the $N=1$ case 
\begin{equation}
\left\{ Q_\alpha ,\bar{Q}_{\dot{\beta}}\right\} =\sigma _{\alpha \dot{\beta}
}^\mu \,p_\mu v_{+}\,\,.  \label{newsusy}
\end{equation}
Here $v_{+}=v_{0^{\prime }}+v_{1^{\prime }}$ is the light-cone component of a
new operator $v_m=i\tilde{\partial}_m$ acting as a momentum in an additional 
$1+1$ dimensional space $y^m=(y^{0^{\prime }},y^{1^{\prime }})$ beyond the
usual four dimensions $x^\mu $. We will see in section-3 that it is possible
to interpret these as the 12th and 11th dimensions respectively. Note that
there are two time coordinates $x^0$ and $y^{0^{\prime }}$; we will show
that in our approach no problems arise due to this fact and that four
dimensional physics looks quite conventional. This superalgebra has an
isometry group $SO(3,1)\otimes SO(1,1)$ which is the direct product of the
Lorentz groups in the $x^\mu $ and $y^m$ spaces. The group $SO(1,1)$
consists of a single parameter corresponding to a boost that mixes the $y^m$
. Its action on the Weyl spinors is an overall scale transformation with
weight $1/2$, while its action on the vector $v_{+}$ is an overall scale
transformation with weight $1$.

In this paper we construct field theory models that provide representations
of this $N=1$ superalgebra. The fields $\Phi (x,y)$ depend on the 4D $x^\mu $
and on the 2D $y^m=y^{\pm }=y^{0^{\prime }}\pm y^{1^{\prime }}.$ We take
derivatives with respect to them $\partial _\mu $ and $\tilde{\partial}_{\pm
}$ such that $p_\mu v_{+}=-\partial _\mu \tilde{\partial}_{+}.$ On this
space the form of the superalgebra is 
\begin{equation}
\left\{ Q_\alpha ,\bar{Q}_{\dot{\beta}}\right\} =-\sigma _{\alpha \dot{\beta}
}^\mu \,\partial _\mu \tilde{\partial}_{+}\,\,.
\end{equation}

If one considers an expansion of the fields in 2D momentum modes 
\begin{equation}
\Phi (x,y)=\sum_{k_{\pm }}\Phi _k(x)\,\,e^{-i\left(
k_{+}y^{+}+k_{-}y^{-}\right) }
\end{equation}
one sees that for each mode the new superalgebra reduces to the standard $
N=1 $ superalgebra, except for a rescaling of the momentum $p^\mu k_{+}$ by
a different amount for each mode. Hence the effect of this type of
``Kaluza-Klein expansion'' on the 4D mass spectrum is very different than
the usual one. We will suggest that the extra space $y^m$ may be related to
family structure (of quarks and leptons) through this new type of expansion.

As explained in section-3 our original motivation for considering this
superalgebra comes from recent developments in S-theory. Some sectors of
S-theory may be described as sectors in which the 12D superalgebra
simplifies to 
\begin{equation}
\left\{ Q_\alpha ,Q_\beta \right\} =\gamma _{\alpha \beta }^{MN}\,\,p_M\,v_N
\label{12pv}
\end{equation}
There is a related version that applies to 13D as well as to
compactifications to lower dimensions ( see \cite{ibstheory}\cite{ibsentropy}
and section-3). The compactified form was recently used to explain the
presence of up to 12 (or 13) hidden dimensions in supersymmetric black holes 
\cite{ibsentropy}. This development provided a strong motivation since the
relevance of the hidden dimensions and of the superalgebra was demonstrated
in a potentially physical system. There were additional hints that such new
superalgebras provide a framework for understanding some deeper structures.
In particular, the 12D version in (\ref{12pv}) with all possible eigenvalues
of the operators $p_M=i\partial /\partial x^M,\,v_N=i\partial /\partial y^N$
that satisfy certain BPS constraints was first suggested in \cite{ibstheory}
as a basis for extending 11D supergravity to 12D supergravity. It was later
used with a fixed eigenvalue of $v_N$ (that breaks $SO(10,2)$ ) in a super
Yang-Mills theory \cite{sezgin}, and in a related matrix model \cite{periwal}
that recasts a matrix version \cite{susskind} of 11D M-theory \cite
{wittenetc} to a 12D version. The same algebra was also understood to be
present in a (2,1) heterotic string \cite{martinec}. From the point of view
of the superalgebra and S-theory the models in \cite{sezgin}\cite{periwal}
\cite{martinec} are incomplete because of the fixed $v_N$. As explained in
section-3, fully $SO(10,2)$ covariant generalizations of these proposals
must exist by allowing all eigenvalues of $v_N\,\,$(as in section-2).
Furthermore such models may be regarded as intermediate steps toward the
construction of the secret theory behind string theory. These provide
additional motivations for studying this type of superalgebra.

In section-3 we will show how the superalgebra (\ref{newsusy}) is embedded
in S-theory and how it is generalized in all possible ways up to 13 higher
dimensions. But at a simpler level one may also consider the $N=1$
superalgebra (\ref{newsusy}) on its own merit as a symmetry structure in
field theory. Thus, a study of its representations in field theory is
undertaken in section-2. In this context we have discovered a new
Kaluza-Klein mechanism for embedding families in higher dimensions. We find
that several families belong together in the same supermultiplet. In
section-2 we provide first examples of representations and family generation
mechanisms that may be generalized in several ways. The fact that this
approach may have phenomenological implications is both surprizing and
welcomed. Connections to M-,F-,S-theories are also given in section-3. In
Section-4 we present the first steps of a superfield formalism and conclude
with some observations.

\section{Field theoretic representations}

The simplest multiplet of standard $N=1$ supersymmetry is the scalar
multiplet that contains the fields ($\phi ,\psi _\alpha ,F)(x)$ that appear
in the Wess-Zumino (WZ) model. For the new $N=1$ superalgebra we will
present four different representations that connect to the WZ representation
upon Kaluza-Klein reduction (there may also be others). We will refer to
them as the scalar-vector multiplets with the fields 
\begin{equation}
(\phi ,\psi _\alpha ,V_{+})(x,y)\quad or\quad (\phi ^{\prime },\psi _\alpha
^{\prime },V_{-}^{\prime })(x,y)
\end{equation}
and the scalar-scalar multiplets with the fields 
\begin{equation}
(\varphi ,\chi _\alpha ,f)(x,y)\quad or\quad (\varphi ^{\prime },\chi
_\alpha ^{\prime },f^{\prime })(x,y).
\end{equation}

The new $N=1$ superalgebra has the isometry $SO(3,1)\times SO(1,1),$
therefore every field must correspond to a representation of this group. Since
we wish to connect to the WZ fields we take a triplet of fields $\,$that are
(scalar, chiral spinor, scalar) under $SO(3,1)$, and next we choose their $
SO(1,1)$ properties as follows. The group $SO(1,1)$ contains a single
parameter corresponding to boosts that mix the ($y^{0^{\prime
}},y^{1^{\prime }}),$ but rescales the light-cone components $y_{\pm }$ with
opposite factors $y_{\pm }\rightarrow \Lambda ^{\pm 1}y_{\pm }$. Under the
boosts the fields $(\phi ,\psi _\alpha ,V_{+})$ undergo scale
transformations with the scale factor raised to the powers $\left(
0,1/2,1\right) $ respectively. The complex conjugates of these fields $(\bar{
\phi},\bar{\psi}_{\dot{\alpha}},\bar{V}_{+})$ have the same $SO(1,1)$
scales. Similarly $(\phi ^{\prime },\psi _\alpha ^{\prime },V_{-}^{\prime })$
and their complex conjugates are assigned $SO(1,1)$ weights $\left(
0,-1/2,-1\right) $, also $(\varphi ,\chi _\alpha ,f)$ have weights $\left(
0,1/2,0\right) $ and $(\varphi ^{\prime },\chi _\alpha ^{\prime },f^{\prime
})$ have weights $\left( 0,-1/2,0\right) .$ The name of the multiplets
``scalar-vector'' refers to the properties of $V_{+},V_{-}^{\prime }$ that
are scalars under $SO(3,1)$ and vectors under $SO(1,1)$. Similarly
``scalar-scalar'' refers to $f,f^{\prime }$ that are scalars under both $
SO(3,1)$ and $SO(1,1)$ .

\subsection{Scalar-vector multiplets}

We guess the following supersymmetry transformation rules for $(\phi ,\psi
_\alpha ,V_{+})$ by imitating the old WZ transformation rules and by
requiring consistency with the isometries $SO(3,1)\times SO(1,1)$

\begin{eqnarray}
\delta \phi =-\varepsilon ^{\prime \alpha }\psi _\alpha ,\,\,\delta \psi
_\alpha =\partial _\mu \tilde{\partial}_{+}\phi \,\sigma _{\alpha \dot{\beta}
}^\mu \bar{\varepsilon}^{\prime \dot{\beta}}+\tilde{\partial}_{+}\tilde{
\partial}_{-}V_{+}\varepsilon _\alpha ,\,\,\delta V_{+}=\bar{\varepsilon}_{
\dot{\beta}}\bar{\sigma}_\mu ^{\dot{\beta}\alpha }\partial ^\mu \tilde{
\partial}_{+}\psi _\alpha   \label{trone} \\
\delta \bar{\phi} = -\bar{\psi}_{\dot{\alpha}}\bar{\varepsilon}^{\prime 
\dot{\alpha}},\,\,\delta \bar{\psi}_{\dot{\alpha}}=\varepsilon ^{\prime
\beta }\sigma _{\beta \dot{\alpha}}^\mu \partial _\mu \tilde{\partial}_{+}
\bar{\phi}\,+\tilde{\partial}_{+}\tilde{\partial}_{-}\bar{V}_{+}\bar{
\varepsilon}_{\dot{\alpha}},\,\,\delta \bar{V}_{+}=\partial ^\mu \tilde{
\partial}_{+}\bar{\psi}_{\dot{\alpha}}\,\bar{\sigma}_\mu ^{\dot{\alpha}\beta
}\varepsilon _\beta   \nonumber
\end{eqnarray}
The fermionic parameters $\varepsilon ^{\prime },\varepsilon $ mix the pairs
($\phi ,\psi )$ and ($\psi ,V_{+})$ respectively. We will build Lagrangians
that are invariant under this transformation for arbitrary global fermionic
parameters $\varepsilon ^{\prime },\varepsilon .$ However for the closure of
the algebra we find that $\varepsilon ^{\prime },\varepsilon $ must be
related as given below.

Taking hermitian conjugation (bar) turns the index $\alpha $ into a dotted
index $\dot{\alpha}$, and also interchanges the order of anticommuting
variables. Applying $C=i\sigma _2$ raises or lowers the index. We have used
it in the following definitions 
\begin{eqnarray}
\sigma _{\alpha \dot{\beta}}^\mu &\equiv &(1,\vec{\sigma}),\quad \left( \bar{
\sigma}^\mu {}\right) ^{\dot{\alpha}\beta }\equiv C(1,\vec{\sigma}
^{*})C^T=\left( 1,-\vec{\sigma}\right) ,\quad \\
\left( \sigma ^\mu \right) ^{\dagger } &=&\sigma ^\mu ,\quad \left( \bar{
\sigma}^\mu \right) ^{\dagger }=\bar{\sigma}^\mu ,\quad \sigma ^\mu \bar{
\sigma}^\nu +\sigma ^\nu \bar{\sigma}^\mu =2\eta ^{\mu \nu }.
\end{eqnarray}
Since we have anticommuting variables, the following rules apply 
\begin{equation}
\varepsilon ^{\prime \alpha }\psi _\alpha =\varepsilon _\beta ^{\prime
}C^{\beta \alpha }\psi _\alpha =\psi _\alpha C^{\alpha \beta }\varepsilon
_\beta ^{\prime }=\psi ^\beta \varepsilon _\beta ^{\prime }=-\psi _\alpha
\varepsilon ^{\prime \alpha }.
\end{equation}
Then the transformation rules are consistent with hermitian conjugation.

By applying two infinitesimal transformations (\ref{trone}) and
antisymmetrizing $\left[ \delta _1,\delta _2\right] \phi $, $\left[ \delta
_1,\delta _2\right] \psi _\alpha ,$ $\left[ \delta _1,\delta _2\right] V_{+}$
we find the closure of the algebra by demanding consistency with eq.(\ref
{newsusy}) 
\begin{eqnarray}
\left[ \delta _1,\delta _2\right] \phi &=&-\varepsilon _2^{\prime \alpha
}\left( \delta _1\psi _\alpha \right) -\left( 1\longleftrightarrow 2\right) 
\nonumber \\
&=&-\varepsilon _2^{\prime \alpha }\left( \partial _\mu \tilde{\partial}
_{+}\phi \,\sigma _{\alpha \dot{\beta}}^\mu \bar{\varepsilon}_1^{\prime \dot{
\beta}}+(\tilde{\partial}_{+}\tilde{\partial}_{-}V_{+})\varepsilon _{1\alpha
}\right) -\left( 1\longleftrightarrow 2\right)  \nonumber \\
&=&\left( \varepsilon _1^{\prime \alpha }\sigma _{\alpha \dot{\beta}}^\mu 
\bar{\varepsilon}_2^{\prime \dot{\beta}}-\varepsilon _2^{\prime \alpha
}\sigma _{\alpha \dot{\beta}}^\mu \bar{\varepsilon}_1^{\prime \dot{\beta}
}\right) \partial _\mu \tilde{\partial}_{+}\phi \,  \label{phi}
\end{eqnarray}
The $V_{+}$ term drops out provided we take 
\begin{equation}
\varepsilon _\alpha ^{\prime }=A\varepsilon _\alpha  \label{phase}
\end{equation}
where $A$ is any complex number. Similarly 
\begin{eqnarray}
\left[ \delta _1,\delta _2\right] V_{+} &=&\bar{\varepsilon}_{2\dot{\beta}}
\bar{\sigma}_\mu ^{\dot{\beta}\alpha }\partial ^\mu \tilde{\partial}
_{+}\left( \partial _\nu \tilde{\partial}_{+}\phi \,\sigma _{\alpha \dot{
\gamma}}^\nu \bar{\varepsilon}_1^{\prime \dot{\gamma}}+\tilde{\partial}_{+}
\tilde{\partial}_{-}V_{+}\varepsilon _{1\alpha }\right) -\left(
1\longleftrightarrow 2\right)  \nonumber \\
&=&\bar{\varepsilon}_{2\dot{\beta}}\bar{\varepsilon}_1^{\prime \dot{\beta}
}\partial ^\mu \partial _\mu \tilde{\partial}_{+}^2\phi +\bar{\varepsilon}_{2
\dot{\beta}}\bar{\sigma}_\mu ^{\dot{\beta}\alpha }\varepsilon _{1\alpha }\,
\tilde{\partial}_{-}\tilde{\partial}_{+}^2\partial ^\mu V_{+}-\left(
1\longleftrightarrow 2\right)  \nonumber \\
&=&\left( \bar{\varepsilon}_{2\dot{\beta}}\bar{\sigma}_\mu ^{\dot{\beta}
\alpha }\varepsilon _{1\alpha }\,-\bar{\varepsilon}_{1\dot{\beta}}\bar{\sigma
}_\mu ^{\dot{\beta}\alpha }\varepsilon _{2\alpha }\,\right) \tilde{\partial}
_{-}\tilde{\partial}_{+}^2\partial ^\mu V_{+}
\end{eqnarray}
Notice that 
\begin{eqnarray}
\varepsilon _2^{\prime \alpha }\sigma _{\alpha \dot{\beta}}^\mu \bar{
\varepsilon}_1^{\prime \dot{\beta}} &=&-\bar{\varepsilon}_1^{\prime \dot{
\beta}}\sigma _{\dot{\beta}\alpha }^{T\mu }\varepsilon _2^{\prime \alpha }=-
\bar{\varepsilon}_{1\dot{\gamma}}^{\prime }C^{\dot{\gamma}\dot{\beta}}\sigma
_{\dot{\beta}\alpha }^{T\mu }\varepsilon _{2\delta }^{\prime }C^{\delta
\alpha }  \nonumber \\
&=&-\bar{\varepsilon}_{1\dot{\gamma}}^{\prime }\left( C\sigma ^{T\mu
}C^T\right) ^{\dot{\gamma}\delta }\varepsilon _{2\delta }^{\prime }=-\bar{
\varepsilon}_{1\dot{\gamma}}^{\prime }\left( \bar{\sigma}^\mu \right) ^{\dot{
\gamma}\delta }\varepsilon _{2\delta }^{\prime }
\end{eqnarray}
Therefore, if we use (\ref{phase}) we obtain 
\begin{equation}
\varepsilon _2^{\prime \alpha }\sigma _{\alpha \dot{\beta}}^\mu \bar{
\varepsilon}_1^{\prime \dot{\beta}}=-\bar{\varepsilon}_{1\dot{\gamma}}\left( 
\bar{\sigma}^\mu \right) ^{\dot{\gamma}\delta }\varepsilon _{2\delta
}\,\,\left| A\right| ^2
\end{equation}
which gives the form 
\begin{equation}
\left[ \delta _1,\delta _2\right] V_{+}=\left( \varepsilon _1^{\prime \beta
}\sigma _{\beta \dot{\alpha}}^\mu \bar{\varepsilon}_2^{\prime \dot{\alpha}
}-\varepsilon _2^{\prime \beta }\sigma _{\beta \dot{\alpha}}^\mu \bar{
\varepsilon}_1^{\prime \dot{\alpha}}\right) \left( -\frac 1{\left| A\right|
^2}\tilde{\partial}_{+}\tilde{\partial}_{-}\right) \partial ^\mu \tilde {
\partial}_{+}V_{+}
\end{equation}
In order to have the same closure property as in (\ref{phi}) we must require 
\begin{equation}
\left( \tilde{\partial}_{+}\tilde{\partial}_{-}+\left| A\right| ^2\right)
V_{+}=0.
\end{equation}
The field is on shell in the extra dimensions but is{\it \ not} on shell
from the point of view of 4D (for the scalar-scalar representation presented
below the field is fully off-shell). We will see that this restriction
follows from a Lagrangian that is fully invariant without using any mass
shell conditions. Next consider the spinor 
\begin{eqnarray}
\left[ \delta _1,\delta _2\right] \psi _\alpha &=&\partial _\mu \tilde{
\partial}_{+}\left( \delta _1\phi \right) \,\sigma _{\alpha \dot{\beta}}^\mu 
\bar{\varepsilon}_2^{\prime \dot{\beta}}+\tilde{\partial}_{+}\tilde{\partial}
_{-}\left( \delta _1V_{+}\right) \varepsilon _{2\alpha }-\left(
1\longleftrightarrow 2\right)  \nonumber \\
&=&-\partial _\mu \tilde{\partial}_{+}\left( \varepsilon _1^{\prime \gamma
}\psi _\gamma \right) \,\sigma _{\alpha \dot{\beta}}^\mu \bar{\varepsilon}
_2^{\prime \dot{\beta}}+\tilde{\partial}_{+}\tilde{\partial}_{-}\left( \bar{
\varepsilon}_{1\dot{\beta}}\bar{\sigma}_\mu ^{\dot{\beta}\gamma }\partial
^\mu \tilde{\partial}_{+}\psi _\gamma \right) \varepsilon _{2\alpha }-\left(
1\longleftrightarrow 2\right)  \nonumber \\
&=&\frac 12\varepsilon _1^{\prime }\sigma _\nu \bar{\varepsilon}_2^{\prime
}\,\,\partial _\mu \tilde{\partial}_{+}\left( \sigma ^\mu \bar{\sigma}^\nu
\right) \psi -\frac 12\left( \bar{\varepsilon}_1\bar{\sigma}_\nu \varepsilon
_2\right) \,\tilde{\partial}_{+}^2\tilde{\partial}_{-}\partial _\mu \left(
\sigma ^\nu \bar{\sigma}^\mu \right) \psi -\left( 1\longleftrightarrow
2\right)  \nonumber \\
&=&\frac 12\left( \varepsilon _1^{\prime }\sigma _\nu \bar{\varepsilon}
_2^{\prime }\,-\varepsilon _2^{\prime }\sigma _\nu \bar{\varepsilon}
_1^{\prime }\right) \left( \sigma ^\mu \bar{\sigma}^\nu +\sigma ^\nu \bar{
\sigma}^\mu \left( -\frac 1{\left| A\right| ^2}\tilde{\partial}_{+}\tilde{
\partial}_{-}\right) \right) \partial _\mu \tilde{\partial}_{+}\psi
\end{eqnarray}
Note that we got an extra minus sign from the interchange of fermions from
line 2 to line 3. The two terms combine to give the desired closure 
\begin{equation}
\left[ \delta _1,\delta _2\right] \psi _\alpha =\left( \varepsilon
_2^{\prime }\sigma _\nu \bar{\varepsilon}_1^{\prime }\,-\varepsilon
_1^{\prime }\sigma _\nu \bar{\varepsilon}_2^{\prime }\right) \partial ^\nu 
\tilde{\partial}_{+}\psi \,_\alpha \,\,\,,
\end{equation}
provided the spinor satisfies 
\begin{equation}
\left( \tilde{\partial}_{+}\tilde{\partial}_{-}+\left| A\right| ^2\right)
\psi _\alpha =0.
\end{equation}
Since both $\psi _\alpha $ and $V_{+}$ are restricted, $\phi $ must also be
restricted by 
\begin{equation}
\left( \tilde{\partial}_{+}\tilde{\partial}_{-}+\left| A\right| ^2\right)
\phi =0,
\end{equation}
in order to have consistent transformation properties. As we will see in
section-3, solutions of a generalized BPS condition involves a time-like
condition on $v^2$. The condition $\,v^2=-\tilde{
\partial}_{+}\tilde{\partial}_{-}=\left| A\right| ^2$ is consistent with
this case, and it is interesting that it followed from the closure of the
superalgebra. As we will see, this is just what we need in order to connect
correctly to standard physics.

So far there is no mass shell condition or equation of motion required in
4D. Therefore these are the properties of the scalar-vector representation $
(\phi ,\psi _\alpha ,V_{+})$ independent of any dynamics. Note that $\left|
A\right| $ plays the role of a label for the representation (like a Casimir
eigenvalue). Next we consider a Lagrangian.

\subsubsection{Free supersymmetric Lagrangian}

The {\it free} Lagrangian we propose is $\pounds _0+\pounds _0^{\prime }$ 
\begin{eqnarray}
\pounds _0 &=&\partial _\mu \tilde{\partial}_{+}\bar{\phi}\,\,\partial ^\mu 
\tilde{\partial}_{-}\phi +\bar{\psi}_{\dot{\alpha}}\bar{\sigma}_\mu ^{\dot{
\alpha}\beta }{}\partial _\mu \tilde{\partial}_{-}\psi _\beta +\tilde{
\partial}_{-}\bar{V}_{+}\,\tilde{\partial}_{-}V_{+},  \nonumber \\
\pounds _0^{\prime } &=&\phi ^{\prime }\left( \tilde{\partial}_{+}\tilde{
\partial}_{-}+\left| A\right| ^2\right) \phi +\psi ^{\prime \alpha }\left( 
\tilde{\partial}_{+}\tilde{\partial}_{-}+\left| A\right| ^2\right) \psi
_\alpha  \\
&&+V_{-}^{\prime }\left( \tilde{\partial}_{+}\tilde{\partial}_{-}+\left|
A\right| ^2\right) V_{+}+h.c.  \nonumber
\end{eqnarray}
$\pounds _0$ contains the original fields $(\phi ,\psi _\alpha ,V_{+})$ and
is hermitian up to total derivatives. $\pounds _0^{\prime }$ is included to
impose the constraints through the equations of motion of the Lagrange
multipliers $(\phi ^{\prime },\psi ^{\prime \alpha },V_{-}^{\prime })$.
Their $SO(1,1)$ weights have to be the opposite of the original fields
because of invariance under $SO(1,1).$ This Lagrangian is unconventional
because of the number of derivatives applied on the fields, and because of
the two time coordinates. According to old wisdom one should expect problems
with ghosts. However one should note that in either the $x$-space or the  $y$
-space there are at the most two derivatives. Also the $y$-space will be 
compactified in a Kaluza-Klein approach. We will see below that this
structure leads to conventional physics in four dimensions without any
problems, i.e. there are no ghosts in the spectrum of this model.   

Each term of the Lagrangian is separately invariant under the
transformations of $(\phi ,\psi _\alpha ,V_{+})$ and of $(\phi ^{\prime
},\psi ^{\prime \alpha },V_{-}^{\prime })$ (given below) for arbitrary
global parameters $\varepsilon ^{\prime },\varepsilon ,$ {\it without using
any constraints}. The constraints that are needed to close the algebra
follow as an equation of motion of the auxiliary fields. Thus, before using
the constraints there is an even larger supersymmetry. We demonstrate the
larger symmetry by applying the supersymmetry transformations on $\pounds _0$
\begin{eqnarray}
\delta \pounds _0 &=&\left( \partial _\mu \tilde{\partial}_{+}\delta \bar{
\phi}\right) \,\partial ^\mu \tilde{\partial}_{-}\phi +\bar{\psi}_{\dot{
\alpha}}\bar{\sigma}_\mu ^{\dot{\alpha}\beta }{}\partial ^\mu \tilde{\partial
}_{-}\left( \delta \psi _\beta \right) +\left( \tilde{\partial}_{-}\delta 
\bar{V}_{+}\right) \,\tilde{\partial}_{-}V_{+}  \nonumber \\
&&+\partial _\mu \tilde{\partial}_{+}\bar{\phi}\,\,\left( \partial ^\mu 
\tilde{\partial}_{-}\delta \phi \right) +\delta \bar{\psi}_{\dot{\alpha}}
\bar{\sigma}_\mu ^{\dot{\alpha}\beta }\partial ^\mu \tilde{\partial}_{-}\psi
_\beta +\tilde{\partial}_{-}\bar{V}_{+}\tilde{\partial}_{-}\left( \delta
V_{+}\right)
\end{eqnarray}
Substituting from (\ref{trone}) we have for the first line 
\begin{eqnarray}
&&-\partial _\mu \tilde{\partial}_{+}\bar{\psi}_{\dot{\alpha}}\bar{
\varepsilon}^{\prime \dot{\alpha}}\,\,\partial ^\mu \tilde{\partial}_{-}\phi
+\bar{\psi}_{\dot{\alpha}}\bar{\sigma}_\mu ^{\dot{\alpha}\beta }{}\partial
^\mu \tilde{\partial}_{-}\left( \partial _\nu \tilde{\partial}_{+}\phi
\,\sigma _{\beta \gamma }^\nu \bar{\varepsilon}^{\prime \dot{\gamma}}+\tilde{
\partial}_{+}\tilde{\partial}_{-}V_{+}\varepsilon _\beta \right)  \nonumber
\\
&&+\tilde{\partial}_{-}\left( \partial ^\mu \tilde{\partial}_{+}\bar{\psi}_{
\dot{\alpha}}\,\bar{\sigma}_\mu ^{\dot{\alpha}\beta }\varepsilon _\beta
\right) \tilde{\partial}_{-}V_{+}\,\,,
\end{eqnarray}
which is a total derivative (without using the constraints) 
\begin{eqnarray}
&&\partial ^\mu \left( \bar{\psi}_{\dot{\alpha}}\bar{\varepsilon}^{\prime 
\dot{\alpha}}\,\tilde{\partial}_{+}\tilde{\partial}_{-}\partial _\mu \phi
\,\,+\bar{\psi}_{\dot{\alpha}}\,\bar{\sigma}_\mu ^{\dot{\alpha}\beta
}\varepsilon _\beta \,\,\tilde{\partial}_{+}\tilde{\partial}
_{-}^2V_{+}\,\right)  \nonumber \\
&&-\tilde{\partial}_{+}\left( \partial _\mu \bar{\psi}_{\dot{\alpha}}\bar{
\varepsilon}^{\prime \dot{\alpha}}\,\partial ^\mu \tilde{\partial}_{-}\phi
+\partial ^\mu \bar{\psi}_{\dot{\alpha}}\,\bar{\sigma}_\mu ^{\dot{\alpha}
\beta }\varepsilon _\beta \,\,\tilde{\partial}_{-}^2V_{+}\right) \\
&&+\tilde{\partial}_{-}\left( \tilde{\partial}_{-}V_{+}\,\,\tilde{\partial}
_{+}\partial ^\mu \bar{\psi}_{\dot{\alpha}}\,\bar{\sigma}_\mu ^{\dot{\alpha}
\beta }\varepsilon _\beta \right) .  \nonumber
\end{eqnarray}
Similarly, the second line gives 
\begin{eqnarray}
&&\partial _\mu \tilde{\partial}_{+}\bar{\phi}\,\,\partial ^\mu \tilde{
\partial}_{-}\left( -\varepsilon ^{^{\prime }\alpha }\psi _\alpha \right) +
\tilde{\partial}_{-}\bar{V}_{+}\,\left( \bar{\varepsilon}_{\dot{\beta}}\bar{
\sigma}_\mu ^{\dot{\beta}\alpha }\partial ^\mu \tilde{\partial}_{+}\tilde{
\partial}_{-}\psi _\alpha \right)  \nonumber \\
&&\,\,+\left( \varepsilon ^{\prime \gamma }\sigma _{\gamma \dot{\alpha}}^\nu
\partial _\nu \tilde{\partial}_{+}\bar{\phi}+\tilde{\partial}_{+}\tilde{
\partial}_{-}\bar{V}_{+}\,\bar{\varepsilon}_{\dot{\alpha}}\right) \,\bar{
\sigma}_\mu ^{\dot{\alpha}\beta }\partial ^\mu \tilde{\partial}_{-}\psi
_\beta
\end{eqnarray}
which is also a total derivative 
\begin{equation}
\partial ^\mu \left( \partial ^\nu \tilde{\partial}_{+}\bar{\phi}
\,\,\varepsilon ^{\prime }\sigma _{\nu \mu }\tilde{\partial}_{-}\psi
\,\,\right) +\tilde{\partial}_{+}\left( \tilde{\partial}_{-}\bar{V}_{+}\,
\bar{\varepsilon}_{\dot{\alpha}}\bar{\sigma}_\mu ^{\dot{\alpha}\beta
}\partial ^\mu \tilde{\partial}_{-}\psi _\beta \right) .
\end{equation}

Now, we turn to $\pounds _0^{\prime }.$ Its variation under the
supertransformation gives 
\begin{eqnarray}
\delta \pounds _0^{\prime } &=&\delta \phi ^{\prime }\left( \tilde{\partial}
_{+}\tilde{\partial}_{-}+\left| A\right| ^2\right) \phi +\delta \psi
^{\prime \alpha }\left( \tilde{\partial}_{+}\tilde{\partial}_{-}+\left|
A\right| ^2\right) \psi _\alpha +\delta V_{-}^{\prime }\left( \tilde{\partial
}_{+}\tilde{\partial}_{-}+\left| A\right| ^2\right) V_{+}  \nonumber \\
&&+\phi ^{\prime }\left( \tilde{\partial}_{+}\tilde{\partial}_{-}+\left|
A\right| ^2\right) \left( -\varepsilon ^{\prime \alpha }\psi _\alpha \right)
+V_{-}^{\prime }\left( \tilde{\partial}_{+}\tilde{\partial}_{-}+\left|
A\right| ^2\right) \left( \bar{\varepsilon}_{\dot{\beta}}\bar{\sigma}_\mu ^{
\dot{\beta}\alpha }\partial ^\mu \tilde{\partial}_{+}\psi _\alpha \right) 
\nonumber \\
&&+\psi ^{\prime \alpha }\left( \tilde{\partial}_{+}\tilde{\partial}
_{-}+\left| A\right| ^2\right) \left( \partial _\mu \tilde{\partial}_{+}\phi
\,\sigma _{\alpha \dot{\beta}}^\mu \bar{\varepsilon}^{\prime \dot{\beta}}+
\tilde{\partial}_{+}\tilde{\partial}_{-}V_{+}\varepsilon _\alpha \right)
+h.c.
\end{eqnarray}
We get $\delta \pounds _0^{\prime }=$total derivative, without using the
constraints, provided the Lagrange multipliers transform under supersymmetry
as follows 
\begin{equation}
\delta \phi ^{\prime }=-\partial _\mu \tilde{\partial}_{+}\psi ^{\prime
\alpha }\sigma _{\alpha \dot{\beta}}^\mu \bar{\varepsilon}^{\prime \dot{\beta
}},\,\,\delta \psi ^{\prime \alpha }=\phi ^{\prime }\varepsilon ^{^{\prime
}\alpha }+\partial ^\mu \tilde{\partial}_{+}V_{-}^{\prime }\bar{\varepsilon}
_{\dot{\beta}}\bar{\sigma}_\mu ^{\dot{\beta}\alpha },\,\,\delta
V_{-}^{\prime }=\tilde{\partial}_{+}\tilde{\partial}_{-}\psi ^{\prime \alpha
}\varepsilon _\alpha
\end{equation}

So, the total free Lagrangian is supersymmetric for arbitrary $\varepsilon
,\varepsilon ^{\prime }\,$without using any constraints. This larger
symmetry closes into a larger set of bosonic operators included in S-theory.
We will not discuss the larger symmetry in any detail but one can see its
general structure in section-3. The smaller superalgebra (\ref{newsusy}) is
represented correctly only after we use the constraints 
\begin{equation}
\varepsilon ^{\prime }=A\varepsilon ,\quad \tilde{\partial}_{+}\tilde{
\partial}_{-}+\left| A\right| ^2=0.
\end{equation}

It is possible to write an additional piece in the free supersymmetric
Lagrangian involving only the primed fields. In that case the primed fields
become propagating fields instead of being Lagrange multipliers. This
Lagrangian is invariant without using the constraints provided the number of
derivatives $\tilde{\partial}_m\,$on the fermion $\psi ^{\prime }$ is cubic.
Because of the high derivatives, and because we are interested in
interpreting the primed fields as non-propagating fields, we refrain from
adding this additional term in the present model.

\subsubsection{On mass shell degrees of freedom}

We can now analyze the equations of motion and determine the content of the
degrees of freedom on mass shell 
\begin{eqnarray}
\bar{\sigma}_\mu ^{\dot{\alpha}\beta }{}\partial _\mu \tilde{\partial}
_{-}\psi _\beta  &=&-\left( \tilde{\partial}_{+}\tilde{\partial}_{-}+\left|
A\right| ^2\right) \psi ^{\prime \dot{\alpha}},\quad \tilde{\partial}
_{-}^2V_{+}=-\left( \tilde{\partial}_{+}\tilde{\partial}_{-}+\left| A\right|
^2\right) V_{-}^{\prime }\,\quad  \\
\tilde{\partial}_{+}\tilde{\partial}_{-}\partial ^\mu \partial _\mu \phi 
&=&-\left( \tilde{\partial}_{+}\tilde{\partial}_{-}+\left| A\right|
^2\right) \phi ^{\prime },\quad \left( \tilde{\partial}_{+}\tilde{\partial}
_{-}+\left| A\right| ^2\right) \left[ \phi ,\psi _\alpha ,V_{+}\right] =0,
\end{eqnarray}
The only solutions of the last equation are expressed as a linear
combination of the following complete basis 
\begin{equation}
e^{-i\left( k_{+}y^{+}+k_{-}y^{-}\right) }\left[ \phi _k\left( x\right)
,\psi _{k\alpha }\left( x\right) ,V_{+k}\left( x\right) \right] ,\quad with
{ \thinspace \thinspace \thinspace \thinspace }k_{+}k_{-}=\left|
A\right| ^2.
\end{equation}
In the other equations the primed fields must be in the same basis (to match
the $y^{\pm }$ dependence). However, the operator $\tilde{\partial}_{+}
\tilde{\partial}_{-}+\left| A\right| ^2$ applied on the primed fields
vanishes on this basis. Therefore, the only solutions are of the form 
\begin{equation}
\left( \tilde{\partial}_{+}\tilde{\partial}_{-}+\left| A\right| ^2\right)
\left[ \phi ^{\prime },\psi _\alpha ^{\prime },V_{-}^{\prime }\right]
=0,\,\quad \left( \tilde{\partial}_{+}\tilde{\partial}_{-}+\left| A\right|
^2\right) \left[ \phi ,\psi _\alpha ,V_{+}\right] =0
\end{equation}
which reduce the original equations of motion to massless field equations in
4D since $\tilde{\partial}_{+}\tilde{\partial}_{-}$ or $\tilde{\partial}_{-}$
cannot vanish on these fields, 
\begin{equation}
\partial ^\mu \partial _\mu \phi =0,\quad \bar{\sigma}_\mu ^{\dot{\alpha}
\beta }{}\partial _\mu \psi _\beta =0,\quad V_{+}=0.
\end{equation}
So the modes $\phi _k\left( x\right) ,\psi _{k\alpha }\left( x\right) $ are
ordinary massless bosonic and fermionic fields in 4D, while $V_{+k}\left(
x\right) =0.$ For each $k^m$ one has the degrees of freedom of a scalar
multiplet of ordinary $N=1$ supersymmetry. The Hilbert space constructed
with these degrees of freedom has no ghosts.

\subsubsection{Families and Kaluza-Klein compactification}

The massless modes are labeled by the Kaluza-Klein momenta $k^m$ in 11th and
12th dimensions$.$ Let us assume that these dimensions are compactified so
that the momenta are quantized as follows 
\begin{equation}
k_{\pm }=\frac{n_{\pm }}{R_{\pm }}
\end{equation}
where $n_{\pm }$ are integers. Then we must choose 
\begin{equation}
\left| A\right| ^2=\frac n{R_{+}R_{-}}
\end{equation}
where $n$ is a positive fixed integer, and the integers $n_{\pm }$ must take all
possible values such that 
\begin{equation}
n_{+}n_{-}=n.  \label{int}
\end{equation}
$n$ is a label of the representation. For fixed $n$ the solutions of (\ref
{int}) are given as follows

\begin{eqnarray}
n &=&1:\quad \left( n_{+},n_{-}\right) =\left( 1,1\right)   \nonumber \\
n &=&2:\quad \left( n_{+},n_{-}\right) =\left( 2,1\right) ,\left( 1,2\right) 
\nonumber \\
n &=&3:\quad \left( n_{+},n_{-}\right) =\left( 3,1\right) ,\left( 1,3\right) 
\label{fam} \\
n &=&4:\quad \left( n_{+},n_{-}\right) =\left( 4,1\right) ,\left( 1,4\right)
,\left( 2,2\right)   \nonumber \\
&&etc.  \nonumber
\end{eqnarray}
where we have listed only the positive values, assuming a positivity
condition for both $k_{\pm }.$

We see that the on mass shell physical modes correspond to free fields that
satisfy the operator conditions 
\begin{equation}
p^2=0,\,\,\,\quad v^2=\frac n{R_{+}R_{-}}.  \label{BPS}
\end{equation}
where $n/R_{+}R_{-}$ characterizes the fixed geometry in the compactified
11th and 12th dimensions (see section-3 for an interpretation of these
conditions as generalized BPS constraints). For fixed geometry there are
only a finite number of solutions as determined by $n$. This could be
interpreted as a ``family quantum number''. We have obtained a finite number
of families for fixed $n$ because $k_{\pm }$ are {\it both} quantized.
Without such a quantization the number of solutions of $k_{+}k_{-}=\left|
A\right| ^2$ is infinite. Furthermore if $n_{\pm }$ are integers but $n$ is not an integer there are no solutions at all. With the quantization of all three integers $n, n_{\pm }$ number theory comes to the
rescue to give a finite number of solutions.

The Kaluza-Klein momenta in the new SUSY multiply the usual momenta $
k_{+}p_\mu $, not add. Therefore their effect is similar to the slope
parameter $\alpha ^{\prime }\,\,$of strings ( but unlike $\alpha ^{\prime },$
the factor $k_{+}$ is not necessarily a constant on all fields). For
massless particles their presence does not change the mass, therefore we
just get repetitions of massless particles, i.e. families.

So far the model is non-interacting. We expect that in the presence of
interactions, for example gauge or gravitational interactions, the operators 
$p_\mu ,v_m$ would be replaced by covariant derivatives in the closure of
the superalgebra. The analog of the BPS conditions (\ref{BPS}) would then
become Laplacians and Dirac operators in the presence of interactions. Then
one may consider their solutions in the presence of non-trivial geometries
in $y^m$ space (analogs of Calabi-Yau, etc., but now in a space with
Minkowski signature). Also, as in sections 3 and 4, one may add the other $c$
compactified dimensions and consider $SO(c+1,1)$ instead of $SO(1,1).$
Evidently the geometry is bound to modify the number of permitted solutions
interpreted as families. While there are some similarities between this
embedding of families in the geometries of higher dimensions, the equations
and the mechanisms are different as compared to the more familiar
Kaluza-Klein mechanism. We have seen already from (\ref{fam}) that there are
new possibilities that had not emerged before. This unexpected wealth of
possibilities may have fruitful phenomenogical applications.

\subsection{Scalar-scalar representations}

Instead of the $SO(1,1)$ vector $V_{+}$ of the previous subsection we now
take an $SO(1,1)$ scalar $f$ and consider the supersymmetry transformation
rules of the fields $(\varphi ,\chi _\alpha ,f)$ that are consistent with
the isometries $SO(3,1)\times SO(1,1)$

\begin{eqnarray}
\,\delta \varphi &=&-\varepsilon ^{\prime \alpha }\chi _\alpha ,\,\,\,\delta
\chi _\alpha =\partial _\mu \tilde{\partial}_{+}\varphi \sigma _{\alpha \dot{
\beta}}^\mu \bar{\varepsilon}^{\prime \dot{\beta}}+s\tilde{\partial}
_{+}f\varepsilon _\alpha ,\,\,\,\delta f=\bar{\varepsilon}_{\dot{\beta}}\bar{
\sigma}_\mu ^{\dot{\beta}\alpha }\partial ^\mu \chi _\alpha \\
\delta \bar{\varphi} &=&-\bar{\chi}_{\dot{\alpha}}\bar{\varepsilon}^{\prime 
\dot{\alpha}},\,\,\,\delta \bar{\chi}_{\dot{\alpha}}=\varepsilon ^{\prime
\beta }\sigma _{\beta \dot{\alpha}}^\mu \partial _\mu \tilde{\partial}_{+}
\bar{\varphi}+s^{*}\tilde{\partial}_{+}\bar{f}\bar{\varepsilon}_{\dot{\alpha}
},\,\,\delta \bar{f}=\partial ^\mu \bar{\chi}_{\dot{\alpha}}\,\bar{\sigma}
_\mu ^{\dot{\alpha}\beta }\varepsilon _\beta  \nonumber
\end{eqnarray}
where $s$ is some complex number to be determined. In this case the $SO(1,1)$
weight of $f$ is $0,$ which is to be contrasted to the previous case. Note
again that we have two independent parameters $\varepsilon ^{\prime
},\varepsilon .$ The Lagrangian presented below is invariant under arbitrary 
$\varepsilon ^{\prime },\varepsilon ,s.$ Closure of the algebra will require
a relation between these parameters, but it will not require any mass shell
conditions as shown below.

By applying two infinitesimal transformations (\ref{trone}) and
antisymmetrizing $\left[ \delta _1,\delta _2\right] \varphi $, $\left[
\delta _1,\delta _2\right] \chi _\alpha ,$ $\left[ \delta _1,\delta
_2\right] f$ we find the closure of the algebra consistent with eq.(\ref
{newsusy}) 
\begin{eqnarray}
\left[ \delta _1,\delta _2\right] \varphi &=&-\varepsilon _2^{\prime \alpha
}\left( \delta _1\chi _\alpha \right) -\left( 1\longleftrightarrow 2\right) 
\nonumber \\
&=&-\varepsilon _2^{\prime \alpha }\left( \partial _\mu \tilde{\partial}
_{+}\varphi \,\sigma _{\alpha \dot{\beta}}^\mu \bar{\varepsilon}_1^{\prime 
\dot{\beta}}+s\tilde{\partial}_{+}f\varepsilon _{1\alpha }\right) -\left(
1\longleftrightarrow 2\right) \\
&=&\left( \varepsilon _1^{\prime \alpha }\sigma _{\alpha \dot{\beta}}^\mu 
\bar{\varepsilon}_2^{\prime \dot{\beta}}-\varepsilon _2^{\prime \alpha
}\sigma _{\alpha \dot{\beta}}^\mu \bar{\varepsilon}_1^{\prime \dot{\beta}
}\right) \partial _\mu \tilde{\partial}_{+}\varphi \,  \nonumber
\end{eqnarray}
The $f$ term drops out provided we take 
\begin{equation}
\varepsilon _\alpha ^{\prime }=A\varepsilon _\alpha
\end{equation}
where $A$ is any complex number. Similarly 
\begin{eqnarray}
\left[ \delta _1,\delta _2\right] f &=&\bar{\varepsilon}_{2\dot{\beta}}\bar{
\sigma}_\mu ^{\dot{\beta}\alpha }\partial ^\mu \left( \partial _\nu \tilde{
\partial}_{+}\varphi \,\sigma _{\alpha \dot{\gamma}}^\nu \bar{\varepsilon}
_1^{\prime \dot{\gamma}}+s\tilde{\partial}_{+}f\varepsilon _{1\alpha
}\right) -\left( 1\longleftrightarrow 2\right)  \nonumber \\
&=&\bar{\varepsilon}_{2\dot{\beta}}\bar{\varepsilon}_1^{\prime \dot{\beta}
}\partial ^\mu \partial _\mu \tilde{\partial}_{+}\varphi +s\bar{\varepsilon}
_{2\dot{\beta}}\bar{\sigma}_\mu ^{\dot{\beta}\alpha }\varepsilon _{1\alpha
}\,\tilde{\partial}_{+}\partial ^\mu f-\left( 1\longleftrightarrow 2\right) 
\nonumber \\
&=&s\left( \bar{\varepsilon}_{2\dot{\beta}}\bar{\sigma}_\mu ^{\dot{\beta}
\alpha }\varepsilon _{1\alpha }\,-\bar{\varepsilon}_{1\dot{\beta}}\bar{\sigma
}_\mu ^{\dot{\beta}\alpha }\varepsilon _{2\alpha }\,\right) \tilde{\partial}
_{+}\partial ^\mu f  \nonumber \\
&=&-\frac s{\left| A\right| ^2}\left( \varepsilon _1^{\prime \beta }\sigma
_{\beta \dot{\alpha}}^\mu \bar{\varepsilon}_2^{\prime \dot{\alpha}
}-\varepsilon _2^{\prime \beta }\sigma _{\beta \dot{\alpha}}^\mu \bar{
\varepsilon}_1^{\prime \dot{\alpha}}\right) \partial ^\mu \tilde{\partial}
_{+}f
\end{eqnarray}
In order to have the same closure property as in (\ref{phi}) we must require 
\begin{equation}
s=-\left| A\right| ^2.
\end{equation}
Next consider the spinor 
\begin{eqnarray}
\left[ \delta _1,\delta _2\right] \chi _\alpha &=&\partial _\mu \tilde{
\partial}_{+}\left( \delta _1\varphi \right) \,\sigma _{\alpha \dot{\beta}
}^\mu \bar{\varepsilon}_2^{\prime \dot{\beta}}+s\tilde{\partial}_{+}\left(
\delta _1f\right) \varepsilon _{2\alpha }-\left( 1\longleftrightarrow
2\right)  \nonumber \\
&=&-\partial _\mu \tilde{\partial}_{+}\left( \varepsilon _1^{\prime \gamma
}\chi _\gamma \right) \,\sigma _{\alpha \dot{\beta}}^\mu \bar{\varepsilon}
_2^{\prime \dot{\beta}}+s\left( \bar{\varepsilon}_{1\dot{\beta}}\bar{\sigma}
_\mu ^{\dot{\beta}\gamma }\partial ^\mu \tilde{\partial}_{+}\chi _\gamma
\right) \varepsilon _{2\alpha }-\left( 1\longleftrightarrow 2\right) 
\nonumber \\
&=&\frac 12\varepsilon _1^{\prime }\sigma _\nu \bar{\varepsilon}_2^{\prime
}\,\,\partial _\mu \tilde{\partial}_{+}\left( \sigma ^\mu \bar{\sigma}^\nu
\right) \chi -\frac s2\left( \bar{\varepsilon}_1\bar{\sigma}_\nu \varepsilon
_2\right) \,\tilde{\partial}_{+}\partial _\mu \left( \sigma ^\nu \bar{\sigma}
^\mu \right) \chi -\left( 1\longleftrightarrow 2\right)  \nonumber \\
&=&\frac 12\left( \varepsilon _1^{\prime }\sigma _\nu \bar{\varepsilon}
_2^{\prime }\,-\varepsilon _2^{\prime }\sigma _\nu \bar{\varepsilon}
_1^{\prime }\right) \left( \sigma ^\mu \bar{\sigma}^\nu +\sigma ^\nu \bar{
\sigma}^\mu \left( -\frac s{\left| A\right| ^2}\right) \right) \partial _\mu 
\tilde{\partial}_{+}\chi
\end{eqnarray}
The two terms combine to give the desired closure provided we use again $
s=-\left| A\right| ^2.$ 
\begin{equation}
\left[ \delta _1,\delta _2\right] \chi _\alpha =\left( \varepsilon
_2^{\prime }\sigma _\nu \bar{\varepsilon}_1^{\prime }\,-\varepsilon
_1^{\prime }\sigma _\nu \bar{\varepsilon}_2^{\prime }\right) \partial ^\nu 
\tilde{\partial}_{+}\chi \,_\alpha \,\,\,.
\end{equation}
In this version we did not need to impose any mass shell constraints in
order to close the algebra.

\subsubsection{Free supersymmetric Lagrangian}

The {\it free} Lagrangian we start with is $\pounds _1$ 
\begin{equation}
\pounds _1=\partial _\mu \tilde{\partial}_{+}\bar{\varphi}\,\,\partial ^\mu 
\tilde{\partial}_{-}\varphi +\bar{\chi}_{\dot{\alpha}}\bar{\sigma}_\mu ^{
\dot{\alpha}\beta }{}\partial _\mu \tilde{\partial}_{-}\chi _\beta -s\tilde{
\partial}_{+}\bar{f}\,\tilde{\partial}_{-}f
\end{equation}
Applying the supersymmetry transformations on $\pounds _0$ we have 
\begin{equation}
\delta \pounds _1=total\,\,\,derivative,  \nonumber
\end{equation}
for any $\varepsilon ^{\prime },\varepsilon ,s.$ The equations of motion
that follow from this free Lagrangian require $v^2f\equiv -\tilde{\partial}
_{+}\tilde{\partial}_{-}f=0$ and $p^2v^2=0$ on both $\varphi ,\chi .$ There
are two classes of solutions 
\begin{eqnarray}
(i) &:&p^2=0,\quad \quad v^2\neq 0, \\
(ii) &:&p^2\neq 0,\quad \quad v^2=0.  \nonumber
\end{eqnarray}
Neither class is physically satisfactory. In class (i) the solution for $f$
is zero while the other fields are massless. This seems fine, but since $v^2$
is not determined there are an infinite number of massless families from the
point of view of 4D. In class (ii) $p^2$ is not required to be on shell by
the equations of motion. To avoid these problems we add an additional part
to the free Lagrangian to enforce the constraint $v^2=M^2\geq 0$ via
Lagrange multipliers 
\begin{equation}
\pounds _{12}=\varphi ^{\prime }\left( \tilde{\partial}_{+}\tilde{\partial}
_{-}+M^2\right) \varphi +\chi ^{\prime \alpha }\left( \tilde{\partial}_{+}
\tilde{\partial}_{-}+M^2\right) \chi _\alpha +f^{\prime }\left( \tilde{
\partial}_{+}\tilde{\partial}_{-}+M^2\right) f+h.c
\end{equation}
Now, we turn to check the invariance of $\pounds _{12}^{\prime }.$ Its
variation under supertransformation gives 
\begin{eqnarray}
\delta \pounds _{12} &=&\delta \varphi ^{\prime }\left( \tilde{\partial}_{+}
\tilde{\partial}_{-}+M^2\right) \varphi +\delta \chi ^{\prime \alpha }\left( 
\tilde{\partial}_{+}\tilde{\partial}_{-}+M^2\right) \chi _\alpha +\delta
f^{\prime }\left( \tilde{\partial}_{+}\tilde{\partial}_{-}+M^2\right) f 
\nonumber \\
&&+\varphi ^{\prime }\left( \tilde{\partial}_{+}\tilde{\partial}
_{-}+M^2\right) \left( -\varepsilon ^{\prime \alpha }\chi _\alpha \right)
+f^{\prime }\left( \tilde{\partial}_{+}\tilde{\partial}_{-}+M^2\right)
\left( \bar{\varepsilon}_{\dot{\beta}}\bar{\sigma}_\mu ^{\dot{\beta}\alpha
}\partial ^\mu \chi _\alpha \right)   \nonumber \\
&&+\chi ^{\prime \alpha }\left( \tilde{\partial}_{+}\tilde{\partial}
_{-}+M^2\right) \left( \partial _\mu \tilde{\partial}_{+}\varphi \,\sigma
_{\alpha \dot{\beta}}^\mu \bar{\varepsilon}^{\prime \dot{\beta}}+s\tilde{
\partial}_{+}f\varepsilon _\alpha \right) +h.c.
\end{eqnarray}
We get $\delta \pounds _{12}^{\prime }=$total derivative, without using the
constraints, provided the primed fields transform under supersymmetry as
follows 
\begin{equation}
\delta \varphi ^{\prime }=-\partial _\mu \tilde{\partial}_{+}\chi ^{\prime
\alpha }\,\sigma _{\alpha \dot{\beta}}^\mu \bar{\varepsilon}^{\prime \dot{
\beta}},\quad \delta \chi ^{\prime \alpha }=\varepsilon ^{\prime \alpha
}\varphi ^{\prime }+\bar{\varepsilon}_{\dot{\beta}}\bar{\sigma}_\mu ^{\dot{
\beta}\alpha }\partial ^\mu f^{\prime },\quad \delta f^{\prime }=s\tilde{
\partial}_{+}\chi ^{\prime \alpha }\varepsilon _\alpha 
\end{equation}
Remarkably, this transformation closes $\left[ \delta _1,\delta _2\right]
(\varphi ^{\prime },\chi ^{\prime },f^{\prime })$ as desired without
requiring any mass shell constraints. So, the total free Lagrangian is
supersymmetric, and the supersymmetry algebra closes as desired, provided $
\varepsilon ^{\prime }=A\varepsilon ,$ \thinspace and $s=-M^2$ as before.
The free model with 
\begin{equation}
\pounds _0^{\left( 1\right) }=\pounds _1+\pounds _{12}
\end{equation}
has a physically satisfactory 4D mass spectrum. Class (ii) is completely
eliminated while in class (i) there are a finite number of families as in
the scalar-vector model of the previous subsection provided
\begin{equation}
M^2=\frac n{R_{+}R_{-}},
\end{equation}
and $n$ is a positive integer.

\subsubsection{scalar-scalar hyper-multiplet}

We could have stopped here, but we also wish to build more models by
exploring the possibility of adding a supersymmetric free Lagrangian
involving only the primed fields. There is one that satisfies $\delta 
\pounds _2=$total derivative: 
\begin{equation}
\pounds _2=\varphi ^{\prime }\bar{\varphi}^{\prime }+\chi ^{\prime \alpha
}\sigma _{\alpha \dot{\beta}}^\mu \partial _\mu \tilde{\partial}_{+}\bar{\chi
}^{\prime \dot{\beta}}-\frac 1s\partial _\mu f^{\prime }\partial ^\mu \bar{f}
^{\prime }.
\end{equation}
By itself the spectrum of this Lagrangian has some of the problems of $
\pounds _1.$ However, when added to the previous terms it becomes
interesting. Each term in the following total Lagrangian is supersymmetric
separately 
\begin{equation}
\pounds _0^{(2)}=\pounds _1+\gamma \pounds _{12}+\pounds _2.
\end{equation}
Note that there is an additional parameter $\gamma $ that cannot be absorbed
into normalizations of the fields. In this model there are two scalar-scalar
representations that are coupled to each other in a supersymmetric invariant
way. The significance of this coupling is that now there are two propagating
fermions $\chi ^\alpha ,\chi ^{\prime \alpha }$ both of which are left
handed $SO(3,1)$ spinors $\left( 1/2,0\right) $ but they have opposite $
SO(1,1)$ chiralities (or weights $\pm 1/2$). Together they are equivalent to
a full Dirac spinor of $SO(1,1)$ as well as of $SO(3,1)$, and their mixing
term in $\gamma \pounds _{12}$ is the analog of a fermion mass term with
mass $m\sim \gamma (\tilde{\partial}_{+}\tilde{\partial}_{-}+M^2)$ from the
point of view of 4D. This interpretation is better understood by analyzing
the coupled equations. One now finds that $\varphi ^{\prime },f$ are
completely solved in terms of the other fields and the remaining two complex
bosons and two chiral fermions form a massive hyper-multiplet of ordinary
supersymmetry in 4D. These remaining fields can be expanded in Kaluza-Klein
modes, where the modes have quantized momenta labeled by $\left(
n_{+},n_{-}\right) $. Each mode satisfies the following mass shell condition 
\begin{eqnarray}
p^2 &=&\frac{\gamma ^2}{k^2}\left( k^2-M^2\right) ^2 \\
&=&\frac{\gamma ^2}{R_{+}R_{-}}\frac{\left( n_{+}n_{-}-n\right) ^2}{
n_{+}n_{-}}
\end{eqnarray}
where we have used $k^2=\frac{n_{+}n_{-}}{R_{+}R_{-}}$ and $M^2=\frac
n{R_{+}R_{-}}.$ Unlike our previous examples, here $k^2$ or the product $
n_{+}n_{-}$ is not fixed. A plot of $p^2$ versus $n_{+}n_{-}$ shows the
following effects. For $n_{+}n_{-}=n$ there are massless modes, $p^2=0,$
provided $n$ is an integer, and this fixed integer (which is a label of the
representation) determines the number of massless families as in (\ref{fam}
). In addition, there are an infinite number of massive Kaluza-Klein modes
for $n_{+}n_{-}\neq n$ such that their 4D mass gets bigger for $n_{+}n_{-}$
increasing toward +infinity as well as for $n_{+}n_{-}$ decreasing toward
zero. There is a mass gap away from from zero mass, $n_{+}n_{-}=n,$ since $
p^2$ has quantized values in units of $\frac{\gamma ^2}{R_{+}R_{-}}$. These
features are compatible with a physical interpretation of the spectrum in
4D. However this model is not entirely satisfactory because the spectrum of $
p^2$ contains tachyons when $n_{+}n_{-}$ is negative ( $k^2\sim $
space-like). Additional input is needed to prevent $n_{+}n_{-}$ from being
negative.   Perhaps interactions, or an appropriate interpretation of the
extra $y$-space in terms of  $p$-branes,  will suggest how to impose $
k^2\geq 0.$ This issue does not arise in the other models presented in this
paper because for them $k^2$ is fixed and positive.

\section{S-theory origins and generalizations}

In this section we describe the algebra (\ref{newsusy}) in the context of a
more general framework in order to display its connections to a secret
theory behind string theory, and to provide a basis for generalizations.

The goal in S-theory \cite{ibstheory} is to extract information about the
secret theory behind string theory by combining the representation structure
of a generalized superalgebra with other information that may be available
about the secret theory through some of its limits such as string theory,
p-brane theory, D-branes and the likes. This strategy is similar to the one
used in the 1960's, with symmetries and current algebras on the one hand and
experimental input on the other, which eventually led to the discovery of
the Standard Model.

S-theory has two types of superalgebras with 32 real supergenerators and 528
real bosonic generators: the $SO(10,2)$ covariant type-A in 12 dimensions
and the $SO(9,1)\times SO(2,1)$ covariant type-B in 13 dimensions. By a
change of basis the same superalgebras may be rewritten in bases that
display other symmetry structures. From the point of view of 10 dimensions
the 32$_{A,B}$ spinors correspond to two 16-component spinors, such that for
the type-A the 10D-chiralities are opposite while for type-B the
10D-chiralities are the same, as in type-A and type-B string theories. The
two types may be embedded in a 13D superalgebra by considering the
64-component spinor space of $SO(11,2).$ Then two different $A,B$
projections reduce the 64-component spinor into distinguishable 32$_{A,B}$
fermions, and those pick out the sets 528$_{A,B}$ out of the $\frac
1264\times 65=78+286+1716$ bosons classified as antisymmetric tensors with
2,3,6 indices under $SO(11,2)$. The A and B types are T-dual to each other
such that T-duality mixes the 13th dimension with the others\footnote{
In the complete theory probably only the self T-dual subset are actually the
same operators while the remainder are T-dual without being identical
operators.}. Therefore, even though there is no $SO(11,2)$ covariant
formalism, thanks to T-duality of string theory we already know that there
is a sense in which all 13 dimensions are connected to each other in the
complete secret theory.

We remind the reader that one cannot consider more than 32 real supercharges 
{\it in the flat limit} of the secret theory. If there were more than 32,
they would show up in 4D as more than $N=8$ supersymmetries, and this is not
permitted by the absence of massless interacting particles with helicities
higher than 2, in the flat limit. Similarly, with 32$_{A,B}$ supercharges
there cannot be more than 528$_{A,B}$ bosonic generators since 528 is the
number of independent components of a $32\times 32$ symmetric matrix.
Special forms of these superalgebras are obtained in representations in
which some of the 528$_{A,B}$ bosons or some of the 32$_{A,B}$ fermions
vanish. The basic hypothesis of S-theory is that in the complete secret
theory all of these operators are realized non-trivially when all of its
sectors are taken into account. In the curved version of the secret theory
more supercharges may exist, but they should vanish as the curvature
vanishes. In considering curved spaces one is interested in what happens to
the superalgebra of the 32$_{A,B}$ supercharges that survive in the flat
limit. Those can close only on the same set of 528$_{A,B}$ bosons, but some
of the latter, as well as some of the 32$_{A,B}$ fermions, could satisfy non
Abelian commutation relations depending on the nature of the curved space.
As suggested in \cite{ibstheory} various curved space models may be
described as contractions of supergroups such as $OSp(1/32),\,OSp(1/64)$ and
other non-Abelian supergroups.

All sectors of the secret theory would fall into some representation of
S-theory. Such sectors include well known theories such as super Yang-Mills,
supergravity and superstring theory. Furthermore M- and F- theories can be
viewed in the same light. This is because the type-A superalgebra contains
the superalgebra of 11D M-theory \cite{wittenetc}, while the type-B
superalgebra contains the superalgebra of 12D F-theory \cite{vafa}, so these
theories could be embedded in a larger theory in 12D and 13D respectively.
Various new compactifications \cite{vafanew} of the secret theory also seem
to be consistent with the overall 12D or 13D algebraic structure of S-theory
(Abelian and non-Abelian). Constructing simple explicit models that provide
representations of the generalized superalgebra of S-theory is likely to
shed more light on the dynamical structure of the secret theory behind
string theory. Section-2 is a small step in this direction and it should
provide an example of the idea expressed in this paragraph.

\subsection{Type-A}

Starting with the type-A superalgebra that contains a 2-brane and a
self-dual 6-brane in 12D \footnote{
The 12D momentum operator $P_M\gamma _{\alpha \beta }^M$ cannot appear, and $
Z_{M_1M_2}$ is not the 12D Lorentz generator. So, this algebra is not the extension of the conformal superalgebra in 12D. The $Z$'s have to do with p-brane open boundaries in flat and curved dimensions, or with wrappings of p-branes in dimensions with non-trivial topologies. The embedding of 11D in 12D
with this interpretation was presented in 1995 in
a conference \cite{ibjapan} as the first suggestion of 12 dimensions as a
step beyond the 11D M-theory. Since this superalgebra is type-A, not type-B,
this 12D is distinguishable than the one suggested later in F-theory \cite{vafa}.
Also, in the type-B superalgebra, one must distinguish the explicit isometry 
$SO(2,1)$ that acts on 3D (including the 13th dimension) from the $SL(2)$ of
U-duality that is not an explicit isometry of the superalgebra, but is used
in describing a 12D F-theory.} 
\begin{eqnarray}
&&\left\{ {Q_\alpha ,Q_\beta }\right\} =\left( S_A\right) _{\alpha \beta } 
\nonumber \\
&&S_A=\frac{1+\gamma _{13}}2C\left( \gamma ^{M_1M_2}\,\,Z_{M_1M_2}\,+\gamma
^{M_1\cdots M_6}\,\,\,Z_{M_1\cdots M_6}^{+}\right)   \label{typea}
\end{eqnarray}
and then reducing to 4 dimensions, one obtains the generalized $N=8$
superalgebra in 4D in a particular basis, extended with all possible 528
bosonic generators \cite{ibsentropy} 
\begin{eqnarray}
\left\{ Q_{\alpha a},Q_{\beta b}\right\}  &=&\left( i\sigma _2\right)
_{\alpha \beta }\,z_{ab}+\left( i\sigma _2\vec{\sigma}\right) _{\alpha \beta
}\,\cdot \vec{F}_{ab}  \nonumber \\
\left\{ \bar{Q}_{\dot{\alpha}\dot{a}},\bar{Q}_{\dot{\beta}\dot{b}}\right\} 
&=&\left( i\sigma _2\right) _{\dot{\alpha}\dot{\beta}}\,z_{\dot{a}\dot{b}
}^{*}+\left( i\sigma _2\vec{\sigma}\right) _{\dot{\alpha}\dot{\beta}}\,\cdot 
\vec{F}_{\dot{a}\dot{b}}^{*}  \label{4d} \\
\left\{ Q_{\alpha a},\bar{Q}_{\dot{\beta}\dot{b}}\right\}  &=&\sigma
_{\alpha \dot{\beta}}^\mu \,\left( \gamma _{a\dot{b}}^m\,P_{\mu m}+\gamma _{a
\dot{b}}^{m_1m_2m_3}A_{\mu m_1m_2m_3}\right)   \nonumber
\end{eqnarray}
where $\mu ,m$ are Lorentz indices for $SO(3,1),\,SO(c+1,1)$ respectively, $
c=6$ is the number of compactified string dimensions, and the extra $(1,1)$
correspond to the 11th and 12th dimensions. The pair of spinor indices $
\alpha a,\dot{\alpha}\dot{a}$ are Weyl spinor indices for the spacetime $
SO(3,1)\,$and internal $SO(c+1,1)$ groups, such that the Weyl projection is
simultaneously left-handed or simultaneously right handed for both indices
(because of the 12D Weyl projection $1+\gamma _{13}$). The ordinary $N=8$
supersymmetry with all of its Lorentz scalar central extensions is obtained
for $c=6$ by keeping only $z_{ab},z_{\dot{a}\dot{b}}^{*},P_{\mu 0^{\prime }},
$ and setting the remaining Lorentz non-scalar operators to zero. In that
sector one may transform to a basis with an $SU(8)$ symmetry, such that the
momentum $P_{\mu 0^{\prime }}$ is a singlet under the $SU(8)$. This $SU(8)$
is the maximal compact group of $E_{7,7}$ of U-duality \cite{julia}. The isometry group $
SO(c+1,1)=SO(7,1)$ is not in this $SU(8)$ or even in the $E_{7,7}$ because $
P_{\mu 0^{\prime }}$ is not a singlet under $SO(c+1,1).$ The web of these
symmetries is described further in \cite{ibjapan}\cite{ibstheory} and it has
been used to explain how the black hole entropy in 4D and 5D contains
information up to 12 (or 13) hidden dimensions \cite{ibsentropy}.

The form of the superalgebra given above may be taken with other values of $
c $, as we will do below in order to study simpler systems with fewer
supersymmetries. In particular $c=0,1$ contains $N=1,2$ supersymmetry.

\subsection{some sectors}

S-theory suggests that the other operators beyond $z_{ab},z_{\dot{a}\dot{b}
}^{*},P_{\mu 0}$ (i.e. the Lorentz non-scalars) also play a role in the
secret theory. Hence we are interested in exploring models that provide
representations of the more general algebra even if they correspond to a
simplified sector of the algebra in which some of the operators vanish, as
long as some of the novel features that relate to the Lorentz non-scalars
are included. With this in mind, a greatly truncated version of the 12D
type-A superalgebra (\ref{typea}) was first suggested in \cite{ibstheory} by
taking $Z_{M_1M_2}=\frac 12(p_{M_1}v_{M_2}-p_{M_2}v_{M_1})$ and $
Z_{M_1\cdots M_6}^{+}=0$ 
\begin{equation}
\left\{ Q_\alpha ,Q_\beta \right\} =\gamma _{\alpha \beta }^{MN}\,\,p_Mv_N
{\ .}\,  \label{newsuper12}
\end{equation}
The generic representations of this superalgebra are long supermultiplets of
minimum dimension $2^{32/2}$  with $2^{15}$ bosons and $2^{15}$ fermions.
Shorter multiplets also exist provided one imposes the generalized BPS
constraint 
\begin{equation}
\det \left( \gamma _{\alpha \beta }^{MN}\,\,p_Mv_N\right) =\left[
p^2v^2-\left( p\cdot v\right) ^2\right] ^{16}=0.
\end{equation}
There are three classes of 12D {\it covariant} solutions of the BPS
constraint: (i) neither $p^2,v^2$ is zero, (ii) one of them is zero, (iii)
both of them are zero. When $p^2$ or $v^2$ are non-zero their signs classify
different representations. The physical signs and solutions must be imposed
through the details of a physical theory. Some such input is the
interpretation of the $Z_{M_1M_2},\,$ $Z_{M_1\cdots M_6}^{+}$ in terms of
p-brane boundaries.  Each solution is distinct in the sense that $SO(10,2)$
transformations cannot relate them, but obviously, solution (i) contains
(ii) and (iii), and solution (ii) contains (iii) as special cases. Examples
of physical representations and the issue of signs were discussed in
section-2 (for the $N=1$ case).

For cases (i) and (ii) there are 16 zero and 16 non-zero supercharges and
the minimal supermultiplet of dimension $2^{16/2}$ contains 128 bosons plus
128 fermions. This is the same set of massless states of 11D membrane theory 
\cite{bsp}, 10D string theory, or 11D supergravity, but in the present case
they are part of the spectrum of a 12D secret theory that contains these
theories. Considering the low energy limit in a field theory context, this
supergravity multiplet would be realized on bi-local fields $\Phi (x^M,y^M)$
on which $p_M,v_M$ act as derivatives $i\partial /\partial x^M,i\partial
/\partial y^M$ respectively. When the BPS constraints are satisfied with 
\begin{equation}
v^2=time-like,\quad v\cdot p=0  \label{timelike}
\end{equation}
and $p^2=0$ on shell, these fields are directly connected to 11D
supergravity fields by a Kaluza-Klein reduction in the $y$-space and keeping
only one eigenvalue of $v_M$. Hence the unreduced theory must be the long
sought $SO(10,2)$ supergravity, as suggested in \cite{ibstheory}. The
non-locality  is a remnant of the extended objects that are needed to
realize this type of superalgebra.

Similarly, for case (iii) there are 24 zero and 8 non-zero supercharges and
the minimal supermultiplet has dimension $2^{8/2}.$ So the Yang-Mills super
multiplet provides a basis for realizing the superalgebra as the $(10,2)$
extension of 10D super Yang-Mills theory. However, such a field theory may
be realized co-covariantlyvariantly provided one uses bi-local fields that satisfy the
constraints 
\begin{equation}
v^2=0,\quad p\cdot v=0  \label{lightlike}
\end{equation}
and take $p^2=0$ on shell.

The superalgebra (\ref{newsuper12}) suggested in \cite{ibstheory} later
found realizations as the supersymmetry algebra for a series of intriguing
models: a 12D Yang-Mills theory \cite{sezgin}, a 12D heterotic $(2,1)$
-string \cite{martinec}, a covariant matrix model \cite{periwal} for a
possible large-N matrix description of M-theory \cite{susskind}. These
models are incomplete from the point of view of representations of the
superalgebra (\ref{newsuper12}) and S-theory. As suggested in \cite
{ibsentropy} to complete the representation space one must take all
eigenvalues of $p_M,v_M$ that are consistent with a given solution of the
constraints (i,ii,iii), rather than taking one of them as a {\it constant}
light-like vector. This requires bi-local fields, as in section-2. Bi-local
fields, that include all the Kaluza-Klein modes consistent with a solution
of the BPS constraints, contain a finite or an infinite number of
eigenvalues of $v_M$ (as in section-2). Only if all such Kaluza-Klein modes
are included can one maintain the 12D covariance. As currently known, the
models in \cite{sezgin}\cite{martinec}\cite{periwal} correspond to keeping
one Kaluza-Klein mode (the constant vector) in an expansion of another
complete theory. 

Another sector of the superalgebra (\ref{4d}) was shown to be relevant for
supersymmetric extremal black holes \cite{ibsentropy}. In this sector one
sets to zero all bosonic operators except for $z_{ab},z_{\dot{a}\dot{b}}^{*}$
and take the special factorized form $P_{\mu m}=p_\mu v_m$. The resulting
superalgebra is covariant under $SO(3,1)\times SO(c+1,1)$ which keeps track
of all 12 dimensions. It was shown that the black hole entropy is invariant
under this isometry and that it contains information about the hidden 12th
or 13th dimension. To do so all eigenvalues of $v_m$ had to be allowed. This
is the first instance in which all eigenvalues of $v_m$ showed up in a
physical system, thus providing encouragement for pursuing this approach
further.

In sections-1,2 of this paper we have considered a sector along the lines of
the last paragraph. In this sector the discussion is simpler, and perhaps
more relevant for possible physical applications. We have also specialized
to the sector of zero central charges since they may be included in later
investigations. Then one has 
\begin{equation}
\left\{ Q_{\alpha a},\bar{Q}_{\dot{\beta}\dot{b}}\right\} =\sigma _{\alpha 
\dot{\beta}}^\mu \,\gamma _{a\dot{b}}^m\,p_\mu v_m  \label{newsuper}
\end{equation}
where $\mu ,m$ are Lorentz indices for $SO(3,1),\,SO(c+1,1)$ respectively as
in (\ref{4d}). In the Weyl sector the gamma matrices may be represented by
hermitian matrices as follows 
\begin{equation}
\sigma ^\mu \,\,p_\mu =p_0+\vec{\sigma}\cdot \vec{p},\quad \gamma
^mv_m=v_{0^{\prime }}+\vec{\gamma}\cdot \vec{v}\,\,,  \label{gammas}
\end{equation}
where in the time directions $\mu =0,m=0^{\prime }$ the Weyl projected gamma
matrices are proportional to unity. The resulting superalgebra may also be
viewed as a reduction of the12D superalgebra of (\ref{newsuper12}) in which
the BPS constraints are satisfied in a sector \footnote{
More generally (\ref{newsuper12}) allows also the components $p_m$ and $
v_\mu .$ In their presence we must have bi-local fields, but we have
specialized to ordinary local fields in 12D by working in the sector in
which $p_m$ and $v_\mu $ are zero. In this way we have lost the full $
SO(10,2)$ but still have an isometry that keeps track of all 12 dimensions.
Case (i) cannot be realized in this sector, but cases (ii) and (iii) remain.}
. In this way with a minimal set of operators one can still probe some novel
sector of S-theory that respects the isometry $SO(3,1)\times \,SO(c+1,1)$.

\subsection{Type-B sectors}

Similarly, one may start from the type-B superalgebra that is covariant
under $SO(9,1)\times SO(2,1)$ \cite{ibstheory} and rewrite it in a 4D basis
by using an explicit covariance $SO(3,1)\times SO(c)\times SO(2,1),$ with $
c=6$. Then the indices on the 32 spinors are $Q_{\alpha Aa},$ with $\alpha
=1,2$, denoting an $SO(3,1)$ spinor, $A=1,2,3,4$ denoting an $SO(6)$ spinor
and $a=1,2$ denoting an $SO(2,1)$ spinor. $\bar{Q}_{\dot{\alpha}\dot{A}a}$
is the hermitian conjugate of the 16 complex $Q_{\alpha Aa}.$ The 10D vector
index is split into a 4D index $\mu $ and a 6D index $j$. Then the 4D, N=8
superalgebra, with 528 real bosonic generators, can be put into the
following form that keeps track of all 13 dimensions labeled by $\mu
=0,1,2,3;\,\,j=1,\cdots ,6;\,m=0^{\prime },1^{\prime },2^{\prime }$ 
\begin{eqnarray}
\left\{ Q_{\alpha Aa},Q_{\beta Bb}\right\} &=&\left( i\sigma _2\right)
_{\alpha \beta }\,\left[ \gamma _{AB}^j\left( i\tau _2\tau ^m\right)
_{ab}\left( P_{jm}+iX_{jm}\right) +\gamma _{AB}^{ijk}\left( i\tau _2\right)
_{ab}\,Y_{ijk}\right]  \nonumber \\
&&+\left( i\sigma _2\vec{\sigma}\right) _{\alpha \beta }\left[ \gamma
_{AB}^i\left( i\tau _2\right) _{ab}\vec{Y}_i+\gamma _{AB}^{ijk}\left( i\tau
_2\tau _m\right) _{ab}\vec{X}_{ijk}^m\right]  \nonumber \\
\left\{ \bar{Q}_{\dot{\alpha}\dot{A}\dot{a}},\bar{Q}_{\dot{\beta}\dot{B}\dot{
b}}\right\} &=&\left( i\sigma _2\right) _{\dot{\alpha}\dot{\beta}}\,\left[
\gamma _{\dot{A}\dot{B}}^j\left( i\tau _2\tau ^m\right) _{ab}\left(
P_{jm}-iX_{jm}\right) +\gamma _{\dot{A}\dot{B}}^{ijk}\left( i\tau _2\right)
_{ab}\,Y_{ijk}^{*}\right]  \nonumber \\
&&+\left( i\sigma _2\vec{\sigma}\right) _{\alpha \beta }\left[ \gamma
_{AB}^i\left( i\tau _2\right) _{ab}\vec{Y}_i+\gamma _{AB}^{ijk}\left( i\tau
_2\tau _m\right) _{ab}\vec{X}_{ijk}^m\right] \\
\left\{ Q_{\alpha Aa},\bar{Q}_{\dot{\beta}\dot{B}\dot{b}}\right\} &=&\sigma
_{\alpha \dot{\beta}}^\mu \,\left( 
\begin{array}{c}
\delta _{A\dot{B}}\,\left( i\tau _2\tau ^m\right) _{ab}P_{\mu m}+\delta _{A
\dot{B}}\,\left( i\tau _2\right) _{ab}\,y_\mu \\ 
+\gamma _{A\dot{B}}^{ij}\,\left( i\tau _2\right) _{ab}Y_{\mu ij}+\gamma _{A
\dot{B}}^{ij}\,\,\left( i\tau _2\tau _m\right) _{ab}\,X_{\mu ij}^m
\end{array}
\right)  \nonumber
\end{eqnarray}
The 528$_B$ bosons labeled by the letters $P,Y,X$ come from a relabeling
of the 528$_B$ bosons that were denoted by the same letters in the $
SO(9,1)\times SO(3,1)$ covariant basis of \cite{ibstheory}. Following the
route of reasoning that led to eq.(\ref{newsuper}), a truncation of this
superalgebra gives the form 
\begin{equation}
\left\{ Q_{\alpha Aa},\bar{Q}_{\dot{\beta}\dot{B}b}\right\} =\sigma _{\alpha 
\dot{\beta}}^\mu \,\delta _{A\dot{B}}\left( i\tau _2\tau ^m\right)
_{ab}\,p_\mu v_m  \label{newsuper3}
\end{equation}
where we may use the $SO(2,1)$ gamma matrices $\tau ^m=\left( -i\tau _2,\tau
_3,-\tau _1\right) $ to get 
\begin{equation}
i\tau _2\tau ^mv_m=v_{0^{\prime }}+\tau _1v_{1^{\prime }}+\tau
_3v_{2^{\prime }}
\end{equation}
which is consistent with the notation in (\ref{gammas}).

\subsection{Reduction to N=1}

Both forms (\ref{newsuper},\ref{newsuper3}) become the standard $N=8$
supersymmetry if one keeps only one eigenvalue of a time-like $v^m,$ since
then it is possible to use the isometry to rotate $v^m$ to the form $\gamma
^mv_m=1.$ Similarly, they reduce to standard $N=4$ supersymmetry if $v^m$ is
light-like and fixed. However, if one allows all possible eigenvalues of a
time-like or light-like $v^m,$ just as all possible eigenvalues of $p^\mu $
are allowed, then the representation space is much richer and includes novel
sectors of $S$-theory. The presence of 12 or 13 dimensions manifests itself
through the two distinct forms (\ref{newsuper},\ref{newsuper3}) in the
corresponding representation spaces. Our purpose is to construct some simple
models that provide explicit representations of this new type of
supersymmetry, with all possible eigenvalues of $v_m,$ as in section-2, with
the hope that such models will shed some light on the dynamics of S-theory.

To begin with, one may start the analysis by neglecting the $c$ compactified
dimensions altogether, and keep only the 11th and 12th dimensions in (\ref
{newsuper}) or the 11th, 12th and 13th dimensions in (\ref{newsuper3}). This
corresponds to setting $c=0$ in eq.(\ref{newsuper},\ref{newsuper3}) and
neglecting the spinor indices that correspond to the $c=6$ dimensions. In
doing so we are really studying a much simpler system with fewer
supersymmetries (as if the supersymmetry has been broken down from $N=8$ to $
N=1)$. With $N=8$ supersymmetry we must have supergravity. To avoid such
complicated systems in the initial stages of this program, we prefer to
start with $N=1$ supersymmetry and slowly work our way to higher $N$. Thus,
in the present paper our goal is to provide some examples of representations
for $N=1$. For higher $N$ one will need to deal with a more complicated set
of auxiliary fields in building up the representations.

In the simplified $c=0$ case, $\gamma _{a\dot{b}}^mv_m$ in (\ref{newsuper})
becomes a $1\times 1$ matrix 
\begin{equation}
\gamma _{a\dot{b}}^mv_m=v_{1^{\prime }}+v_{0^{\prime }}\equiv v_{+}
\end{equation}
where $1^{\prime },0^{\prime }$ represent the 11th and 12th dimensions
respectively. The indices $a,\dot{b}=1$ will be suppressed from now on, and
the new superalgebra will be written in the form given in (\ref{newsusy}).
This connects to the standard $N=1$ supersymmetry if only one Kaluza-Klein
mode of $v_m$ is kept. Similarly in the type-B superalgebra (\ref{newsuper3}
) setting $c=0$ corresponds to neglecting the indices $A,\dot{B}$ and
writing 
\begin{equation}
\left\{ Q_{\alpha a},\bar{Q}_{\dot{\beta}b}\right\} =\sigma _{\alpha \dot{
\beta}}^\mu \,\,p_\mu \,\,\left( v_{0^{\prime }}+\sigma _1v_{1^{\prime
}}+\sigma _3v_{2^{\prime }}\right) _{ab}
\end{equation}
This connects to standard $N=2$ supersymmetry if only one Kaluza-Klein mode
of $v_m$ is kept. The same form follows from the type-A superalgebra (\ref
{newsuper}) if one takes $c=1.$ Then $v_{2^{\prime }}$ represents one of
compactified string dimensions $c$ rather than the 13th dimension. The map
between these two equivalent cases corresponds to T-duality that mixes the
13th dimension with the compactified string dimensions.

\section{superspace}

One may wonder whether the results in section-2 can be recast in a
superfield formalism. The hope is that this would provide the calculus of
representation theory for our superalgebra and help in constructing and
analyzing interacting theories. Here we give a brief summary of such an
attempt, which is an imitation of the ordinary superfield formulation with
some new twists. However our formulation is only partially successful
because the calculus of the representations turns out to be more tricky than
just the superfield formulation. It also requires a non-trivial product rule
for superfields which remains to be constructed.

Consider the superalgebra (\ref{newsuper}) with $SO(3,1)\times SO(c+1,1)$
isometry. Introduce fermionic coordinates $\theta ^{\alpha a}$ and their
hermitian conjugates $\bar{\theta}^{\dot{\alpha}\dot{a}}$classified by the
isometry in one to one correspondence to the supercharges. Then the
following representation of supercharges satisfy the algebra 
\begin{equation}
Q_{\alpha a}=\frac \partial {\partial \theta ^{\alpha a}}-\frac 12\sigma
_{\alpha \dot{\beta}}^\mu \gamma _{a\dot{b}}^m\bar{\theta}^{\dot{\beta}\dot{b
}}\partial _\mu \tilde{\partial}_m,\quad \bar{Q}_{\dot{\beta}\dot{b}}=\frac
\partial {\partial \bar{\theta}^{\dot{\beta}\dot{b}}}-\frac 12\theta
^{\alpha a}\sigma _{\alpha \dot{\beta}}^\mu \gamma _{a\dot{b}}^m\partial
_\mu \tilde{\partial}_m  \label{charges}
\end{equation}
Furthermore we may introduce covariant derivatives that antianticommute with
both of these charges 
\begin{equation}
D_{\alpha a}=\frac \partial {\partial \theta ^{\alpha a}}+\frac 12\sigma
_{\alpha \dot{\beta}}^\mu \gamma _{a\dot{b}}^m\bar{\theta}^{\dot{\beta}\dot{b
}}\partial _\mu \tilde{\partial}_m,\quad \bar{D}_{\dot{\beta}\dot{b}}=\frac
\partial {\partial \bar{\theta}^{\dot{\beta}\dot{b}}}\_+\frac 12\theta
^{\alpha a}\sigma _{\alpha \dot{\beta}}^\mu \gamma _{a\dot{b}}^m\partial
_\mu \tilde{\partial}_m
\end{equation}
A general superfield $\Phi (x,y,\theta ,\bar{\theta})$ is a double
polynomial in powers of $\theta ,\bar{\theta}$ with coefficients that are
ordinary fields that have consistent $SO(3,1)\times SO(c+1,1)$ assignments.
A supersymmetry transformation of all the fields is given as 
\begin{equation}
\delta \Phi (x,y,\theta ,\bar{\theta})=\left( \varepsilon ^{\alpha
a}Q_{\alpha a}+\bar{\varepsilon}^{\alpha a}\bar{Q}_{\alpha a}\right) \Phi
(x,y,\theta ,\bar{\theta})  \label{transf}
\end{equation}
The supersymmetry transformation of the components is read off by comparing
the powers of $\theta ,\bar{\theta}$ on both sides.

As in usual supersymmetry, we introduce the concept of a chiral superfield
defined by 
\begin{equation}
\bar{D}_{\dot{\beta}\dot{b}}\Phi ^{\left( chiral)\right) }(x,y,\theta ,\bar{
\theta})=0
\end{equation}
The solution of this equation is 
\begin{equation}
\Phi ^{\left( chiral)\right) }(x,y,\theta ,\bar{\theta})=\exp \left( \frac
12\theta ^{\alpha a}\sigma _{\alpha \dot{\beta}}^\mu \gamma _{a\dot{b}}^m
\bar{\theta}^{\dot{\beta}\dot{b}}\partial _\mu \tilde{\partial}_m\right)
F\left( x,y,\theta \right)
\end{equation}
where $F\left( x,y,\theta \right) $ is the general polynomial involving only 
$\theta .$ Note that, because of the double derivative, the exponential
factor is not the translation operator on the $x,y$ coordinates.

Now, let's specialize to the $N=1$ case, for which $a,\dot{b}=1$ and
therefore this index is suppressed. The chiral superfield can have at the
most two powers of $\theta .$ If one respects the $SO(1,1)$ assignments one
finds, for example, the scalar-scalar chiral supermultiplet 
\begin{equation}
\Phi ^{\left( chiral)\right) }=\exp \left( \frac 12\theta ^\alpha \sigma
_{\alpha \dot{\beta}}^\mu \bar{\theta}^{\dot{\beta}}\partial _\mu \tilde{
\partial}_{+}\right) \left( \phi +\theta ^\alpha \chi _\alpha +\theta
^\alpha \theta _\alpha \tilde{\partial}_{+}f\right) (x,y).
\end{equation}
The supersymmetry transformation applied as a differential operator in the
form of eq.(\ref{transf}) gives the  transformation rules displayed in
section-2 for the components $(\phi ,\chi _\alpha ,f).$ Closure of the
superalgebra is guaranteed by the construction of eq.(\ref{charges}).
Evidently this property is automatically generalized to higher dimensions by
the superfield formalism. For the purpose of defining (at least some)
representations, as above, the superfield formalism given here is clearly
useful.

One may think that this formulation supplies the technique for writing
interactions. Unfortunately this does not work in a straightforward manner.
If one takes a function of the superfield $W(F)$, e.g. a polynomial, and
constructs a new chiral superfield   
\begin{equation}
\exp \left( \frac 12\theta ^{\alpha a}\sigma _{\alpha \dot{\beta}}^\mu
\gamma _{a\dot{b}}^m\bar{\theta}^{\dot{\beta}\dot{b}}\partial _\mu \tilde{
\partial}_m\right) W(F)
\end{equation}
then the transformation law of the original superfields $F$ are not
compatible with the transformation applied on the  superfield $W(F),$ if
super transformations are applied naively by $\delta =\varepsilon ^{\alpha
a}Q_{\alpha a}+\bar{\varepsilon}^{\alpha a}\bar{Q}_{\alpha a}$ as a
differential operator. This is because $Q_{\alpha a}$ contains the double
derivative structure $\partial _\mu \tilde{\partial}_m$ which is not
distributive as a single derivative structure. That is, it does not satisfy
the Leibnitz rule on naive products of the superfield. Hence representations
are not combined into new irreducible ones by naive superfield
manipulations. The correct combination rules remain to be discovered. This
probably requires the construction of a ``star product'' of superfields that
is compatible with the Leibnitz rule. At this stage it is not clear whether
component methods or superfield methods will be more efficient in providing
the techniques for constructing interactions.

In this paper we concentrated on the simplest $N=1$ superalgebra (\ref
{newsuper}) with only two new dimensions $y^m$ and constructed some of its
representations in the context of field theory. Our purpose was to provide
some concrete examples of representations and to show that they can be
connected to familiar 4D physics. Surprizingly, a new mechanism for
embedding a few families in higher compactified dimensions emerged. The
possibility of phenomenological applications is intriguing and encouraging
for pursuing further the ideas in this paper. The construction of
representations of the higher dimensional cases is aided by the superfield
formalism suggested here.

\section {Acknowledgments}

We would like to thank I. Bakas, S. Ferrara and A. Schwimmer for discussions. The work of I. Bars was partially supported by the U.S. Department of Energy under grant number DE-FG03-84ER40168.  The work of C. Kounnas is partially supported by EEC grant ERB-FMRX-CT 96/0045.


\end{document}